%% file: main.tex
\documentclass[compsoc,conference,a4paper,10pt,times]{IEEEtran}

\usepackage{listings}
\usepackage{xcolor}

\definecolor{codegreen}{rgb}{0,0.6,0}
\definecolor{codegray}{rgb}{0.5,0.5,0.5}
\definecolor{codepurple}{rgb}{0.58,0,0.82}
\definecolor{backcolour}{rgb}{0.95,0.95,0.95}
\definecolor{syclblue}{rgb}{0.0,0.4,0.8}

\lstdefinestyle{syclStyle}{
 language=C++,
 backgroundcolor=\color{backcolour},
 commentstyle=\color{codegreen},
 keywordstyle=\color{syclblue}\bfseries,
 numberstyle=\tiny\color{codegray},
 stringstyle=\color{codepurple},
 basicstyle=\ttfamily\footnotesize,
 breakatwhitespace=false,
 breaklines=true,
 captionpos=b,
 keepspaces=true,
 numbers=left,
 numbersep=3pt,
 showspaces=false,
 showstringspaces=false,
 showtabs=false,
 tabsize=2,
 frame=single,
 frameround=tttt,
 framexleftmargin=8mm,
 xleftmargin=8mm, 
 aboveskip=12pt,
 belowskip=12pt,
 lineskip=1pt
}

\lstdefinelanguage{SYCLCPP}[]{C++}{
    morekeywords={sycl, queue, buffer, handler, range, id, 
                  access, mode, read, write, single_task, 
                  parallel_for, submit, wait, get_access,
                  uint32_t, uint64_t, size_t, inline,
                  TDX, TDX-IO, GPU, RDMA, RoCE},
    morecomment=[l]{//},
    morecomment=[s]{/*}{*/},
    morestring=[b]",
    sensitive=true
}

\lstdefinestyle{enhancedSyclStyle}{
    language=SYCLCPP,
    backgroundcolor=\color{backcolour},   
    commentstyle=\color{codegreen}\itshape,
    keywordstyle=\color{syclblue}\bfseries,
    numberstyle=\tiny\color{codegray},
    stringstyle=\color{codepurple},
    basicstyle=\ttfamily\small,
    breakatwhitespace=false,         
    breaklines=true,                 
    captionpos=b,                    
    keepspaces=true,                 
    numbers=left,                    
    numbersep=10pt,                  
    showspaces=false,                
    showstringspaces=false,
    showtabs=false,                  
    tabsize=4,
    frame=shadowbox,
    rulesepcolor=\color{codegray},
    framexleftmargin=5mm,
    xleftmargin=5mm,
    aboveskip=15pt,
    belowskip=15pt,
    lineskip=1.2pt,
    columns=flexible
}

\lstdefinestyle{cleanStyle}{
    language=C++,
    basicstyle=\ttfamily\small,
    keywordstyle=\color{syclblue}\bfseries,
    commentstyle=\color{codegreen},
    stringstyle=\color{codepurple},
    numbers=none,
    frame=none,
    backgroundcolor=\color{white},
    showspaces=false,
    showstringspaces=false,
    tabsize=4,
    breaklines=true,
    aboveskip=8pt,
    belowskip=8pt
}

\input{preamble}

\begin{document}

\title{Scalable GPU-Based Integrity Verification for Large Machine Learning Models}

\author{\IEEEauthorblockN{Marcin Spoczynski}
\IEEEauthorblockA{\textit{Intel Labs}\\
Hillsboro, Oregon, USA\\
marcin.spoczynski@intel.com}
\and
\IEEEauthorblockN{Marcela S. Melara}
\IEEEauthorblockA{\textit{Intel Labs}\\
Hillsboro, Oregon, USA\\
marcela.melara@intel.com}
}
\maketitle

\input{0abstract}
\input{1introduction}
\input{3problem}
\input{4overview}
\input{5implementation}
\input{6discussion}
\input{7conclusion}

\bibliographystyle{IEEEtran}
\bibliography{references}

\end{document}

%% file: preamble.tex
\usepackage{amsmath,amssymb,amsfonts}
\usepackage{cite}
\usepackage{color}
\usepackage[inline]{enumitem}
\usepackage{graphicx}
\usepackage[breaklinks=true]{hyperref}
\usepackage{microtype}
\usepackage{textcomp}
\usepackage{xcolor}
\usepackage{xspace}
\usepackage[T1]{fontenc}
\usepackage{algorithm}
\usepackage{algpseudocode}
\usepackage{algorithmicx}
\usepackage{graphicx}
\usepackage{textcomp}
\usepackage{xcolor}
\usepackage{listings, listings-rust}
\usepackage{tikz}
\usepackage{booktabs}

\usetikzlibrary{shapes,arrows,positioning}

\newif\ifcommentary
\commentarytrue

\ifcommentary
\newcommand{\todo}[1]{{\color{red}\textbf{TODO:} #1}\xspace}
\newcommand{\fixme}[1]{{\color{red}\textbf{!! FIXME: #1 !!}}}
\newcommand{\msm}[1]{[{\color{magenta}msm: #1}]}

\else
\newcommand{\todo}[1]{}
\newcommand{\fixme}[1]{}
\newcommand{\msm}[1]{}
\fi

\algrenewcommand\algorithmicindent{1.0em}  

%% file: 0abstract.tex
\begin{abstract}
We present a security framework that strengthens distributed machine learning by standardizing integrity protections across CPU and GPU platforms and significantly reducing verification overheads. Our approach co-locates integrity verification directly with large ML model execution on GPU accelerators, resolving the fundamental mismatch between how large ML workloads typically run (primarily on GPUs) and how security verifications traditionally operate (on separate CPU-based processes), delivering both immediate performance benefits and long-term architectural consistency.

By performing cryptographic operations natively on GPUs using dedicated compute units (e.g., Intel Arc's XMX units, NVIDIA's Tensor Cores)~\cite{intel2025xe,nvidia2024h100}, our solution eliminates the potential architectural bottlenecks that could plague traditional CPU-based verification systems when dealing with large models. This approach leverages the same GPU-based high-memory bandwidth and parallel processing primitives that power ML workloads~\cite{perotti2025heterogeneous}, ensuring integrity checks keep pace with model execution---even for massive models exceeding 100GB.

This framework establishes a common integrity verification mechanism that works consistently across different GPU vendors and hardware configurations. By anticipating future capabilities for creating secure channels between trusted execution environments and GPU accelerators, we provide a hardware-agnostic foundation that enterprise teams can deploy regardless of their underlying CPU and GPU infrastructures.
\end{abstract}


%% file: 1introduction.tex
\section{Introduction}\label{sec:introduction}

The exponential growth in machine learning (ML) model complexity has fundamentally transformed the security requirements of AI systems.
Modern large language models (LLMs) now exceed 175 billion parameters~\cite{brown2020language},
with some approaching trillion-parameter scales~\cite{fedus2022switch},
necessitating distributed deployment across multiple GPUs and nodes.

Yet, the emergence of sophisticated model supply chain attacks has highlighted critical gaps in the security of ML model lifecycles~\cite{badnets2019supply}.
Recent incidents involving backdoored models~\cite{gu2019badnets}, poisoned training data~\cite{carlini2023poisoning},
and compromised model repositories~\cite{jiang2022modelhubs} demonstrate the urgent need for efficient model lifecycle integrity verification.

Traditional software signing, while valuable, is insufficient for ML artifacts that require specialized handling due to their size, 
continuous evolution through fine-tuning, and complex dependency graphs spanning datasets, base models, and derived variants.
To ensure model authenticity and detect tampering of model artifacts between different ML lifecycle stages and stakeholders,
a growing number of industry efforts provide tools for ML model signing~\cite{openssf2025modelspec}, provenance~\cite{slsa-for-models2024} and lifecycle transparency~\cite{atlas2025}.

However, these massive models, often exceeding 100 GB in size, present unprecedented challenges for integrity verification, secure deployment, and provenance tracking throughout their entire lifecycle~\cite{bommasani2022opportunities}.
In particular, CPU-based approaches to model verification, while acceptable for small or quantized models~\cite{atlas2025},
can create significant bottlenecks that compromise both performance and security in production large model environments~\cite{wang2024enhancing}.
CPU-based approaches therefore fail to provide the comprehensive measurement and attestation capabilities required for modern ML supply chains~\cite{kreuzberger2023mlops}.

Current model integrity verification systems rely on CPU-mediated cryptographic (hashing) to capture model authenticity and measurements.
However, this approach introduces a fundamental architectural mismatch: while GPUs perform the vast majority of large ML computation,
security operations remain CPU-bound, creating potential bottlenecks and increasing the attack surface.
For a typical 100 GB model deployment, CPU-based SHA-256 verification can require several minutes, while GPU inference operates at much faster timescales.
This disparity not only impacts system boot time and deployment latency but also necessitates complex data movement between CPU and GPU memory spaces,
introducing security vulnerabilities like time-of-check-time-of-use (TOCTOU) attacks.

While proposals for integration of model signing with GPU-accelerated computation have demonstrated significant performance gains over CPU-mediated signing and verification~\cite{gan2025sentry}, they do not currently provide integrity and attestation for the
computation, leaving the model signing process exposed to existing CPU-GPU attack vectors.

To address these issues, we introduce a framework for scalable and hardware-attested model integrity using commodity GPUs.

\subsection{Model Signing and Measurement Challenges}

Current model signing initiatives, including the OpenSSF Model Signing (OMS) specification~\cite{openssf2025modelspec} and emerging industry standards, 
face fundamental scalability and integration challenges when applied to modern AI systems.
The cryptographic measurement of large models presents unique requirements that distinguish ML model integrity from traditional software:

\textbf{Scale-dependent Performance}: Model signing must accommodate artifacts ranging from lightweight quantized models (sub-GB) 
to massive foundation models (1 TB+), requiring scalable cryptographic primitives that maintain security guarantees across this range.
Current CPU-based signing approaches exhibit linear scaling degradation, making real-time signing and verification at every lifecycle stage impractical for production workflows.

\textbf{Lifecycle Complexity}: Unlike traditional software, ML models undergo continuous evolution through fine-tuning, 
quantization, pruning, and adaptation processes. Each transformation requires re-measurement and re-signing, 
creating cascading verification requirements throughout the model lifecycle.

\textbf{Multi-stakeholder Provenance}: ML model deployment typically involves multiple parties: dataset providers, 
foundation model creators, fine-tuning organizations, and deployment operators, each requiring independent provenance attestation capabilities.
Traditional single-signature approaches are inadequate for this multi-party trust model, 
requiring verification frameworks that can track stakeholder contributions while maintaining efficient verification.

\textbf{CPU-mediated Verification}: The integration of model signing with GPU-accelerated computation introduces
additional architectural requirements. Modern ML deployment increasingly relies on GPU-to-GPU direct memory access (GPUDirect),
bypassing CPU involvement for performance optimization.
However, current signing frameworks require CPU-mediated verification, 
breaking the optimized data path and introducing security vulnerabilities through unnecessary data movement.

Our framework addresses these challenges by enabling GPU-based model hashing and signing.
Similar to prior work~\cite{gan2025sentry}, we leverage Merkle tree structures to enable efficient re-verification
of modified model components, while also adding incremental measurement capabilities \emph{across} ML lifecycle stages.

\subsection{Hardware-Attested Model Integrity}

Intel's Trust Domain Extensions (Intel TDX)~\cite{intel2023tdx} provide hardware-based isolation and attestation capabilities~\cite{intel2025tdx} that offer a promising foundation for secure AI systems~\cite{wang2024confidential,chrapek2024fortify}.
Intel TDX enables the creation of virtual machine (VM) granular trusted execution environments where code and data remain isolated from malicious hypervisors and co-located tenants.

Furthermore, Intel TDX Connect specification~\cite{intel2023tdxconnect} implements the TEE Device Interface Security Protocol v1.0 (TDISP)~\cite{pci2022tdisp}, extending these trust boundaries to include accelerator devices
and establish direct secure channels between Intel TDX VMs and GPU hardware.
These hardware features enable a new class of model integrity mechanisms that combine cryptographic measurement with hardware attestation.

Rather than relying solely on software-based signature verification as in OMS,
our framework leverages hardware measurement roots of trust to create unforgeable attestations of model integrity state.

\begin{table}[t]
\centering
\small 
\caption{Framework capabilities: current hardware vs. future Intel TDX Connect}
\setlength{\tabcolsep}{4pt} 
\begin{tabular}{lcc}
\toprule
\textbf{Capability} & \textbf{Now} & \textbf{Intel TDX+} \\
\midrule
GPU-accelerated cryptography & \checkmark & \checkmark \\
Verification speedup & \checkmark & \checkmark \\
PyTorch integration & \checkmark & \checkmark \\
Hardware attestation & -- & \checkmark \\
Tamper-proof verification & -- & \checkmark \\
\bottomrule
\end{tabular}
\label{tab:capabilities}
\vspace{-0.3cm} 
\end{table}

\subsection{GPU-Native Security Architecture}

This paper presents a novel framework that offloads cryptographic operations for ML model integrity verification from CPUs to GPU accelerators,
while addressing critical limitations in current solutions, such as vulnerability to tampering during the verification process itself and the performance
bottlenecks that prevent frequent re-verification throughout the model lifecycle.
Our approach anticipates upcoming Intel TDX Connect capabilities and provides immediate benefits through GPU-accelerated hashing and Merkle tree construction~\cite{rfc6962certificate}.
By co-locating model data and integrity verification operations on the same accelerator,
we eliminate the performance penalties and security vulnerabilities associated with CPU-GPU data movement.
Thus, our work complements frameworks such as Atlas~\cite{atlas2025}, that address critical security concerns in the ML model supply chain
by focusing specifically on the performance and architectural aspects of model integrity verification, while maintaining compatibility with existing supply chain security solutions.

\begin{figure}[t!]
\centering
\includegraphics[width=0.45\textwidth]{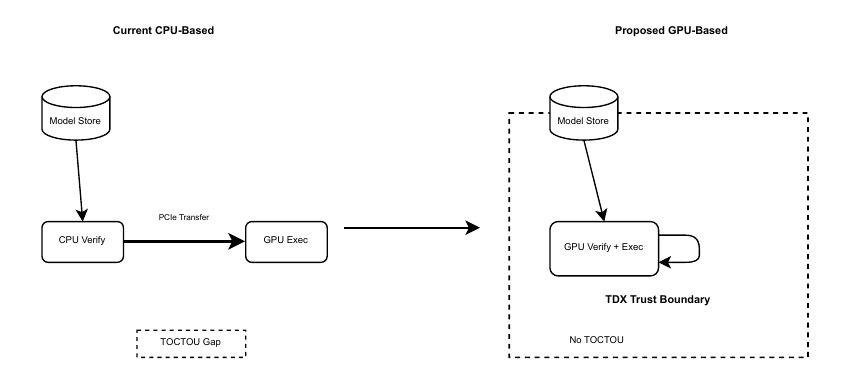}
\caption{Architectural comparison between current CPU-based and proposed GPU-accelerated model integrity verification systems. The current approach (left) suffers from performance bottlenecks due to CPU-mediated verification and CPU-GPU data movement.  The proposed approach (right) co-locates verification and execution within the same GPU platform and Intel TDX trust boundary, eliminating TOCTOU vulnerabilities through unified processing.}
\label{fig:current_vs_proposed_architecture}
\end{figure}

Our framework enables efficient cryptographic model integrity verification that can scale to large model pipelines through four main innovations:

\textbf{Scalable Cryptographic Measurement}: GPU-native hash computation using kernels optimized for parallel execution on modern accelerators~\cite{wang2019gpu} offer performance improvements over CPU implementations while maintaining cryptographic security guarantees.
Our design supports both traditional hash functions (e.g., SHA-256, SHA-384) and emerging post-quantum alternatives,
providing future-proof security for long-lived model deployments.

\textbf{Hierarchical Integrity Verification}: Parallel Merkle tree construction algorithms~\cite{merkle1987digital} enable real-time integrity verification of multi-gigabyte model shards,
supporting both batch and streaming verification modes.
This hierarchical approach enables our framework to check integrity at multiple levels of granularity—individual layers, model components, and full model assemblies, facilitating efficient incremental verification during model updates and fine-tuning operations.

\textbf{Hardware-Software Co-attestation}: Integration of software-based cryptographic verification with hardware attestation capabilities creates unforgeable evidence of model integrity that resists both software and hardware-based attacks.
This co-attestation approach provides defense-in-depth security while enabling new verification capabilities 
such as zero-knowledge model integrity proofs and privacy-preserving model authentication.

\textbf{Runtime Integrity Monitoring}: Trusted execution hardware like Intel TDX performs VM memory integrity checks, ensuring that the model verification process itself occurs within verified, isolated compute environments. This prevents tampering with the verification operation and ensures that integrity measurements are trustworthy. Our framework enables efficient re-verification at multiple pipeline stages, allowing detection of unauthorized modifications between lifecycle phases.

The integration of our GPU-accelerated integrity verification with Intel TDX attestation creates a unified security architecture
where model measurement, signing, attestation, verification and computation can occur within the same trust boundary.
This approach significantly reduces the attack surface compared to traditional CPU-mediated verification schemes,
while simultaneously improving system boot times and deployment efficiency.
Furthermore, our framework enables independent verification of model integrity by multiple parties
without requiring access to proprietary model parameters, enabling trustworthy multi-party model deployment.

\begin{figure}[htbp]
\centering
\includegraphics[width=0.45\textwidth]{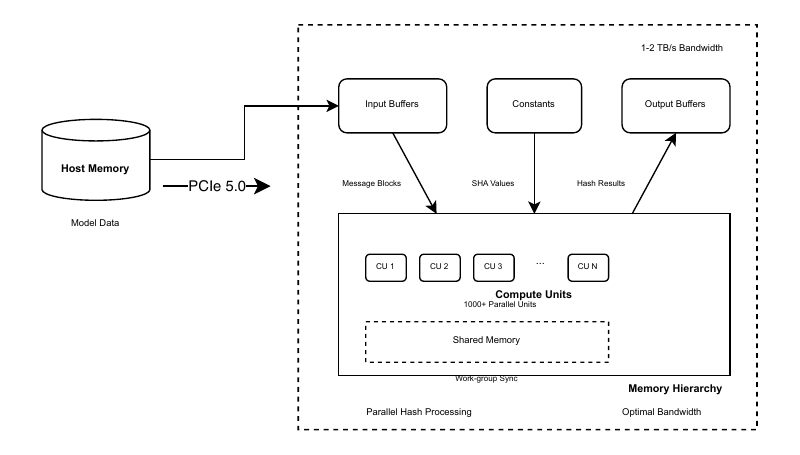}
\caption{GPU memory organization for parallel hash computation showing the hierarchical memory layout that enables efficient SYCL kernel execution. Model parameters are transferred from host memory via PCIe to GPU global memory, where specialized buffers organize data for parallel processing across thousands of compute units. The memory hierarchy includes input buffers for message blocks, constants buffers for SHA round values, output buffers for hash results, and shared memory for work-group coordination, achieving optimal memory bandwidth utilization and parallel throughput.}
\label{fig:gpu_memory_layout}
\end{figure}

\subsection{Production Deployment Considerations}

In addition to providing enhanced attestability for GPU-based model integrity verification,
our approach addresses operational requirements for production ML systems that current solutions fail to meet:

\textbf{Continuous Integration Compatibility}: Seamless integration with existing ML development pipelines, 
supporting automated signing and verification within CI/CD workflows without introducing deployment bottlenecks.
Our GPU-accelerated approach reduces model signing time from minutes to seconds, 
enabling real-time security integration in high-velocity development environments.

\textbf{Multi-environment Portability}: Support for model integrity verification across diverse deployment environments—
from edge devices with integrated GPUs to distributed cloud deployments spanning multiple data centers.
Our SYCL-based implementation provides hardware abstraction while maintaining optimization for specific GPU architectures.

Our GPU-accelerated SHA-256 and SHA-384 implementations demonstrate the potential for significant performance improvements
over CPU implementations for large model verification tasks.

\subsection{Research Contributions}

This paper makes the following specific contributions to GPU-accelerated cryptography for ML model integrity:

\begin{itemize}
    \item \textbf{Scalable Model Measurement Framework}: We develop GPU-native algorithms for hierarchical model integrity measurement and verification,
    enabling real-time integrity checking of multi-gigabyte model shards with sub-millisecond latency per gigabyte of model data.
    Our approach supports incremental measurement for efficient handling of model updates and fine-tuning operations.
    
    \item \textbf{Hardware-Attested Integrity Architecture}: We design a comprehensive framework that integrates GPU-accelerated cryptographic verification
    with hardware attestation capabilities, demonstrating practical deployment in production AI systems with Intel TDX integration.
    
    \item \textbf{Kernels for GPU-Accelerated Cryptography}: We design and implement SYCL-based SHA-256 and SHA-384 kernels
    optimized for parallel execution on modern GPU architectures, achieving significant performance improvements over CPU implementations
    while maintaining full cryptographic compatibility with standard algorithms and emerging post-quantum alternatives.

    \item \textbf{Practical Implementation}: We provide a complete implementation with PyTorch integration, 
    CI/CD pipeline compatibility, and comprehensive evaluation across diverse GPU architectures and model scales.
\end{itemize}

\subsection{Paper Organization}

The remainder of this paper provides a comprehensive treatment of GPU-accelerated model integrity verification:

\textbf{Section~\ref{sec:problem}} establishes the problem context by analyzing current integrity verification challenges in ML systems, presents our threat model with detailed attack vectors, and identifies opportunities for GPU-accelerated solutions that co-locate security operations with model execution.

\textbf{Section~\ref{sec:implementation}} presents our GPU-accelerated integrity verification framework, including Intel Xe-optimized SYCL cryptographic kernels, integration with the Atlas ML lifecycle attestation system, and PyTorch framework integration for production deployments.

\textbf{Section~\ref{sec:discussion}} examines scalability characteristics across large-scale model hierarchies, multi-GPU deployment architectures, version management strategies, and performance analysis at scale.

\textbf{Section~\ref{sec:conclusion}} summarizes our contributions, discusses broader implications for AI security, acknowledges limitations, and outlines future research directions including cryptographic extensions and hardware integration advances.

%% file: 3problem.tex
\section{Problem Statement \& Threat Model}\label{sec:problem}

\subsection{Model Signing and Attestation Challenges}

\subsubsection{Multi-Stakeholder Trust Models}

Modern AI system deployments involves complex trust relationships that traditional software signing approaches cannot adequately address. A typical production scenario may involve:

\begin{itemize}
    \item \textbf{Dataset Providers}: Organizations contributing training data with specific licensing and usage constraints
    \item \textbf{Foundation Model Creators}: Entities developing foundation models that serve as starting points for specialization
    \item \textbf{Fine-tuning Organizations}: Teams adapting foundation models for specific domains or tasks
    \item \textbf{Model Optimizers}: Services providing quantization, pruning, or hardware-specific optimizations
    \item \textbf{Deployment Operators}: Infrastructure providers executing ML pipelines in production environments
    \item \textbf{Regulatory Bodies}: Authorities requiring compliance verification for deployed AI systems
\end{itemize}

Each stakeholder requires independent model integrity attestation capabilities that maintain accountability for individual contributions to the final deployed model.
Traditional single-signature schemes fail to capture these complex trust relationships, necessitating compound signature frameworks that can efficiently verify multi-party attestations.

Current approaches to multi-party signing suffer from combinatorial complexity growth. For a model involving $n$ stakeholders, verification time scales as $O(n)$ for sequential verification or requires complex
coordination protocols for parallel verification. With some production models involving dozens of contributing parties, verification becomes a significant operational bottleneck.

While this work does not implement compound signature verification directly, our GPU-accelerated approach provides the performance foundation necessary to enable efficient verification
in multi-stakeholder environments, where multiple signatures may need to be verified sequentially or in parallel.
\subsubsection{Model Lifecycle Complexity}

ML models undergo continuous evolution that distinguishes them from traditional software artifacts:

\textbf{Incremental Updates}: Fine-tuning operations modify only subsets of model parameters, requiring incremental measurement and re-signing rather than complete re-verification of the entire model.
Current approaches lack efficient incremental attestation capabilities, forcing complete re-measurement for minor parameter updates.

\textbf{Parameter-Efficient Adaptations}: Techniques such as Low-Rank Adaptation (LoRA)~\cite{hu2021lora} and
prefix tuning modify models through additive parameters rather than direct weight updates.
These adaptations require specialized measurement approaches that can verify the integrity of both base models and adaptation parameters while preserving the cryptographic binding between them.

\textbf{Dynamic Model Composition}: Modern deployment may involve real-time composition of multiple models, adapters, and prompt templates.
Current attestation systems cannot attest to a composition of systems while enabling efficient verification of individual components.

\subsubsection{Supply Chain Security Gaps}
Current ML supply chain security approaches fail to address several critical attack vectors:

\textbf{Model Repository Compromise}: Centralized model repositories represent high-value targets for attackers seeking to inject malicious models into downstream systems.
Current approaches rely on repository-level security controls that may be insufficient against sophisticated adversaries with persistent access.

\textbf{Build System Integrity}: The complex toolchains used for model training, optimization, and deployment introduce numerous opportunities for compromise.
Unlike traditional software build systems, ML pipelines involve stochastic processes, dataset dependencies, and hardware-specific optimizations that complicate reproducible build verification.

\textbf{Dependency Chain Attacks}: ML models depend on complex software stacks including framework libraries, hardware drivers, and system dependencies.
Compromise of any dependency can affect model integrity, but current verification approaches focus only on the model artifact itself rather than the complete execution environment.

\textbf{Training Data Provenance}: The integrity of trained models fundamentally depends on the integrity of training data, but current approaches lack efficient mechanisms
for cryptographically binding training data attestations to derived models. This gap enables data poisoning attacks that may not be detectable through model-only verification.

\subsection{Model Integrity Verification Challenges}

Traditional cryptographic verification approaches, designed for smaller software artifacts, 
fail to meet the performance and security requirements of contemporary ML models, while lacking the sophisticated measurement and
attestation capabilities required for complex ML supply chains.

\subsubsection{Scale and Performance Bottlenecks}

For a typical 100 GB model, verification and pipeline execution occur on different processors with distinct performance characteristics and memory systems.
This architectural separation impacts system boot times, auto-scaling responsiveness, and pipeline efficiency.

Additionally, the separation between verification and pipeline execution hardware creates operational challenges.
Thus, whenever model artifacts are verified, data must be transferred between CPU and GPU memory operating with different memory interconnect (i.e., PCIe bus) bandwidth characteristics: CPUs typically provide 100 GB/s while GPUs offer 1 TB/s or more.

This disparity forces AI system developers to choose between security and performance, often resulting in compromised verification schemes or degraded system responsiveness.
For production ML systems requiring continuous model updates through fine-tuning, quantization, or parameter-efficient adaptation techniques,
the cumulative verification overhead becomes prohibitive. A typical production deployment may require verification of dozens of model variants daily,
with current CPU-based approaches consuming hours of computational resources that could otherwise support pipeline workloads.

Co-locating model verification with pipeline execution addresses these challenges by performing security operations on the same hardware.

\subsubsection{Security of CPU-Mediated Verification}

The fundamental design of current model integrity verification introduces several critical limitations.
Data transfers between the CPU and GPU across the PCIe bus not only degrade model run-time performance but also expose an attack surface 
for model shards or metadata to be to be intercepted during transit.
Specifically, this creates time-of-check-time-of-use (TOCTOU) vulnerabilities
whereby an attacker with access to GPU memory or the memory interconnect could alter model artifacts after CPU-mediated verification is complete but before GPU-based pipeline execution begins.

Furthermore, the separation of verification and pipeline execution domains complicates the implementation of continuous integrity monitoring.
That is, real-time integrity checking of the model verification process during different pipeline stages becomes impractical when verification operations must occur on a separate processing unit with limited memory bandwidth to the pipeline execution environment.

\subsection{Limitations of Current Approaches}

\subsubsection{CPU-Based Cryptographic Verification}

Current model integrity verification systems rely on CPU-based cryptographic operations that are fundamentally mismatched to the scale and deployment characteristics of modern AI systems.
Standard approaches involve computing cryptographic hashes (typically SHA-256) of model parameters on the CPU and comparing these hashes to reference values provided by model publishers or signing authorities.

This approach suffers from several critical limitations:

\textbf{Performance Scalability}: CPU-based hash computation exhibits linear scaling with model size, making verification time proportional to model complexity.
For state-of-the-art models approaching 1TB in size, verification times can exceed tens of minutes, making real-time or near-real-time deployment impossible.

\textbf{Memory Bandwidth Constraints}: CPUs have limited memory bandwidth compared to modern GPUs, creating bottlenecks when processing large model datasets.
The memory bandwidth disparity between CPU and GPU architectures (typically 100GB/s vs 1TB/s+) means that CPU-based verification cannot leverage the high-bandwidth memory systems designed for AI workloads.

\textbf{Lack of Parallelization}: Traditional CPU-based hash computation is largely sequential, failing to leverage the massive parallelism available in modern accelerator architectures.
Even with multi-core CPU implementations, the degree of parallelism remains limited compared to GPU architectures with thousands of compute units.

\subsubsection{Inadequate Integration with ML Frameworks}

Existing integrity verification systems operate as external processes or preprocessing steps that are poorly integrated with ML pipelines. This separation creates several operational challenges:

\textbf{Deployment Complexity}: Model verification becomes a separate operational concern that must be managed independently of ML pipelines, increasing deployment complexity and potential points of failure.

\textbf{Resource Utilization Inefficiency}: CPU resources used for verification cannot be leveraged for pipelines, while GPU resources remain idle during verification phases, leading to poor overall resource utilization.

\textbf{Limited Continuous Monitoring}: The separation between verification and pipeline execution hardware makes continuous integrity checking difficult to implement, as the verification process cannot be fully monitored.

\subsubsection{Insufficient Model Signing Capabilities}

Current approaches to ML model signing suffer from fundamental limitations that compromise their effectiveness in production environments:

\textbf{Single-stakeholder Assumption}: Most existing signing frameworks assume a single trusted signer, failing to address the multi-party nature of modern ML development and deployment workflows.

\textbf{Monolithic Model Treatment}: Current approaches treat models as monolithic artifacts, requiring complete re-signing for any parameter updates rather than supporting efficient incremental signing of modified components.

\textbf{Limited Metadata Integration}: Existing frameworks provide limited support for cryptographically binding model metadata (training procedures, data sources, validation results) to the signed model artifact.

\textbf{Weak Lineage Tracking}: Current systems cannot efficiently verify complex model derivation relationships, such as fine-tuning lineages or model merging operations.

\subsection{Security Implications of Current Limitations}

\subsubsection{Extended Attack Windows}

The performance limitations of CPU-based verification create extended time windows during which systems are vulnerable to attack. During the verification phase, models cannot be used in a pipeline,
but the verification process itself may expose sensitive information or create opportunities for side-channel attacks.

Long verification times also necessitate caching of verification results, which introduces additional security concerns.
Cached verification status may become stale if model parameters are modified after verification, and the cache itself becomes a potential target for attack.

\subsubsection{Incomplete Threat Coverage}

Current CPU-based verification systems primarily address supply chain integrity but provide limited protection against runtime attacks.
Once initial verification is complete, continuous integrity monitoring as a model evolves in a pipeline is typically absent or limited, leaving systems vulnerable to runtime modification attacks.


\subsubsection{Accountability Gaps}

Current single-signature approaches create accountability gaps in multi-stakeholder ML deployments. When a model exhibits unexpected behavior,
it becomes difficult to determine which stakeholder's contribution may be responsible, complicating incident response and forensic analysis.

The lack of granular signing capabilities also prevents selective disclosure of model components to different stakeholders, limiting the ability to implement principle-of-least-privilege access controls in collaborative ML development.

\subsection{Threat Model for ML Model Integrity}

We derive the threat model for our GPU-based model integrity framework from the challenges discussed above.

\subsubsection{Trust Assumptions}

Our threat model assumes the following trusted components:
\begin{itemize}
	\item Intel TDX-enabled CPU with attestation capabilities and secure key management
	\item GPU compute units and execution environment within Intel TDX trust boundaries
	\item Future Intel TDX Connect secure device interface enabling GPU attestation
	\item Cryptographic primitives and hash functions (SHA-256, SHA-384, post-quantum alternatives)~\cite{fips2015secure}
	\item Hardware-based root of trust and attestation infrastructure
	\item Secure timestamp services for signature validity periods
	\item Certificate authorities and public key infrastructure for stakeholder authentication
\end{itemize}

We consider the following components as \emph{untrusted} or potentially compromised:
\begin{itemize}
	\item Hypervisor and host operating system
	\item Network infrastructure and model distribution channels
	\item Storage systems and model repositories
	\item Co-located virtual machines or containers
	\item Memory interconnects and PCIe fabric outside Intel TDX boundaries
	\item Model supply chain components and build systems
	\item External training data sources and preprocessing pipelines
	\item Third-party model optimization and conversion services
\end{itemize}

\subsubsection{Attack Vectors Against Model Integrity}

\textbf{Supply Chain Attacks:} may compromise model integrity during training, storage, or distribution phases.
While comprehensive supply chain security is addressed by frameworks like Atlas~\cite{atlas2025}, our GPU-accelerated verification provides the performance necessary to
enable frequent re-verification throughout the model lifecycle, reducing vulnerability windows.

\textbf{Runtime Model Tampering:} Attackers with access to system memory or GPU memory could modify model parameters after initial verification. This category includes:
\begin{enumerate}
	\item \textbf{Memory Corruption Attacks}: Direct modification of model parameters in GPU memory through buffer overflows or memory corruption vulnerabilities
	\item \textbf{Cold Boot Attacks}: Recovery of model parameters from GPU memory after system shutdown, potentially allowing offline analysis and modification
	\item \textbf{DMA Attacks}: Malicious peripheral devices accessing GPU memory through direct memory access, bypassing CPU-based security mechanisms
	\item \textbf{Gradient-based Attacks}: Exploitation of model update mechanisms to gradually modify parameters in ways that evade detection
\end{enumerate}

\textbf{Time-of-Check-Time-of-Use (TOCTOU) Attacks:} The temporal separation between CPU-based verification and GPU-based pipeline execution creates opportunities for race condition attacks.
Adversaries may modify model artifacts during the window between verification completion and pipeline execution.

\begin{figure}[h]
	\centering
	\includegraphics[width=0.45\textwidth]{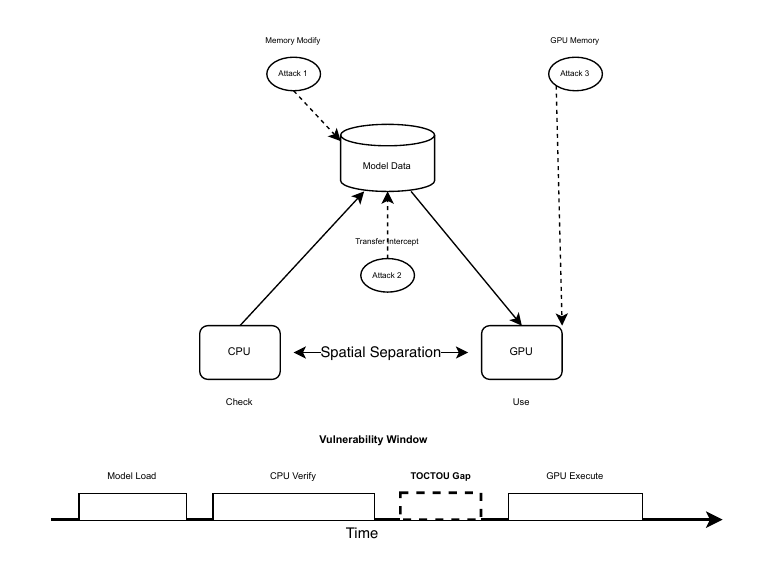}
	\caption{Time-of-Check-Time-of-Use vulnerability in current CPU-mediated model verification systems. The temporal and spatial separation between verification and pipeline execution creates attack opportunities.}
	
	\label{fig:toctou_vulnerability}
\end{figure}

\textbf{Signature and Attestation Attacks:} Sophisticated adversaries may target the cryptographic infrastructure supporting model signing:
\begin{itemize}
	\item \textbf{Key Compromise}: Theft or compromise of signing keys belonging to trusted stakeholders
	\item \textbf{Certificate Authority Attacks}: Compromise of PKI infrastructure to issue fraudulent certificates
	\item \textbf{Timestamp Manipulation}: Attacks on time sources to enable replay or signature validity attacks
	\item \textbf{Multi-party Protocol Attacks}: Exploitation of coordination protocols in compound signature schemes
\end{itemize}

\textbf{Model Loading Attacks:} Sophisticated adversaries may attempt to modify model parameters during the loading phase
between storage and GPU memory, exploiting the gap between deployment-time verification and actual model execution.

\textbf{Attestation Evasion:} Attackers may attempt to circumvent hardware attestation mechanisms through:
\begin{itemize}
	\item \textbf{Firmware Compromise}: Modification of GPU firmware or system BIOS to provide false attestation measurements
	\item \textbf{Hardware Emulation}: Use of modified hardware or virtualization to simulate trusted execution environments
	\item \textbf{Side-channel Attacks}: Extraction of attestation keys or measurements through power analysis, timing attacks, or electromagnetic emissions
\end{itemize}

%% file: 4overview.tex
\section{GPU-Native Model Integrity Overview}

Our GPU-accelerated integrity verification framework addresses several research questions
through a unified security architecture design, compound signature verification algorithms, 
and novel cryptographic kernel implementations.

\subsection{Research Questions and Objectives}

This work addresses the following research questions:

\begin{enumerate}
	\item \textbf{Performance Scalability}: Can GPU-accelerated cryptographic operations provide sufficient performance improvements to enable real-time integrity verification of large-scale AI models and compound signature schemes?
	
	\item \textbf{Security Equivalence}: Do GPU-based integrity verification implementations maintain the same cryptographic security guarantees as CPU-based implementations while providing enhanced attestation capabilities?
	
	\item \textbf{Architectural Integration}: How can GPU-accelerated integrity verification be integrated with existing ML frameworks, signing standards, and emerging hardware security features?
	
	\item \textbf{Practical Deployment}: What are the operational and security implications of deploying GPU-based integrity verification in production AI systems with complex stakeholder relationships?
	
	\item \textbf{Continuous Verification}: Can GPU co-location enable practical continuous integrity monitoring during pipeline execution without significant performance impact?
\end{enumerate}

\subsection{Computational Advantages}

Modern GPUs provide several architectural advantages for cryptographic operations that current verification systems fail to leverage:

\textbf{Massive Parallelism}: GPUs contain thousands of compute units that can process hash computations in parallel, potentially providing order-of-magnitude performance improvements over sequential CPU implementations.

\textbf{High Memory Bandwidth}: GPU memory systems provide significantly higher bandwidth than CPU memory, enabling faster processing of large model datasets without memory bottlenecks.

\textbf{Co-location Benefits}: Performing verification operations on the same hardware that executes pipelines eliminates data movement overhead and reduces TOCTOU attack opportunities.

\subsection{Integration with Hardware Security Features}

The emergence of hardware security features such as Intel TDX and upcoming Intel TDX Connect capabilities creates opportunities for unified security architectures where integrity verification occurs within the same trust boundary as model execution.

Intel TDX provides CPU-based attestation and isolation capabilities that can anchor trust for GPU-based security operations.
TDX Connect extensions will enable secure communication channels between Intel TDX domains and accelerator devices, potentially eliminating the need to trust intermediate system components.

\begin{figure}[h]
	\centering
	\includegraphics[width=0.45\textwidth]{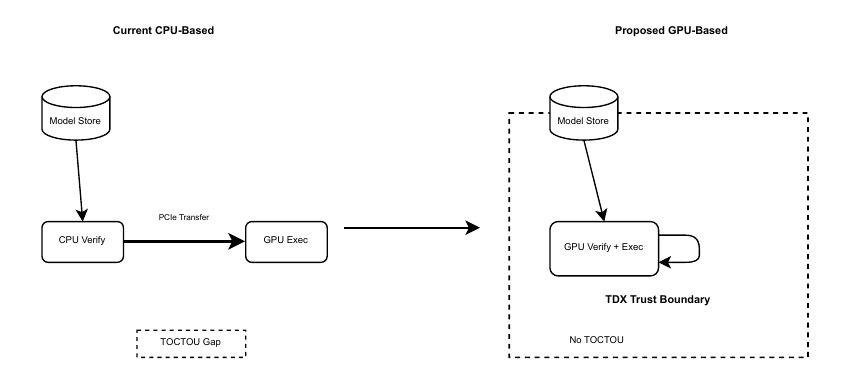}
	\caption{Proposed integration of GPU-accelerated integrity verification with Intel TDX trust boundaries. Upcoming Intel TDX Connect capabilities enable unified trust domains spanning CPU and GPU resources.}
	\label{fig:tdx_integration}
\end{figure}

\subsection{Enhanced Model Signing Capabilities}

GPU acceleration enables several advanced model signing capabilities that are impractical with CPU-based approaches:

\textbf{Real-time Compound Signature Verification}: Parallel processing of multiple signatures enables efficient verification of multi-stakeholder attestations without introducing deployment bottlenecks.

\textbf{Incremental Signature Updates}: GPU-accelerated Merkle tree operations enable efficient re-computation of signatures for incrementally updated models, supporting continuous integration workflows.

\textbf{Privacy-preserving Verification}: GPU acceleration makes zero-knowledge proof verification practical for scenarios where model parameters must remain confidential during verification.

\subsection{Alignment with Industry Standards}

Recent industry initiatives such as the OpenSSF Model Signing specification~\cite{openssf2025modelspec} and
secure model attestation systems~\cite{atlas2025} emphasize the
importance of cryptographic verification throughout the model lifecycle. These frameworks provide standardized interfaces and protocols that can be enhanced through
GPU-accelerated implementations without requiring fundamental changes to existing security policies or procedures.

Our approach extends these frameworks by providing high-performance verification capabilities that can scale to meet the demands of production AI systems while maintaining compatibility with existing security standards and practices.

%% file: 5implementation.tex
\section{Implementation}\label{sec:implementation}

This section details our GPU-accelerated model integrity verification framework, focusing on Intel Xe-optimized SYCL kernels~\cite{intel2025xe} and integration with Rust-based ML lifecycle attestation systems.
Our implementation provides model verification capabilities spanning supply chain attestation to runtime integrity monitoring, with particular emphasis on verification during model loading operations.

\subsection{Intel Xe-Optimized Cryptographic Kernels}

Our implementation leverages Intel's SYCL framework~\cite{sycl2020spec} with specific optimizations for Intel Xe GPU architecture, providing portable high-performance cryptographic operations across
diverse GPU hardware while maximizing performance on Intel Xe and integrated graphics. Implementation details including code listings are provided for reproducibility; detailed source code and build configurations are available in the project repository.

\subsubsection{Architecture-Specific Optimizations}

Intel Xe architecture provides several unique characteristics that enable significant cryptographic performance improvements:

\textbf{64-bit SIMD Processing}: Intel Xe supports efficient 64-bit operations with vectorization capabilities, well-suited for SHA-384/512 algorithms. Our implementation exploits this through custom vectorized data types:

\begin{lstlisting}[language=C++,  style=syclStyle, caption={Intel Xe Vectorized 64-bit Operations}]
struct uint64x2_t {
    uint64_t x, y;
    
    uint64x2_t(uint64_t a, uint64_t b) : x(a), y(b) {}
    
    uint64x2_t operator+(const uint64x2_t& other) const {
        return uint64x2_t(x + other.x, y + other.y);
    }
    
    uint64x2_t operator^(const uint64x2_t& other) const {
        return uint64x2_t(x ^ other.x, y ^ other.y);
    }
    
    uint64x2_t operator&(const uint64x2_t& other) const {
        return uint64x2_t(x & other.x, y & other.y);
    }
};
\end{lstlisting}

\textbf{Sub-group Collaborative Processing}: Intel Xe sub-groups can enable efficient collaborative computation across related threads in suitable workloads. Our SHA-384 implementation uses sub-group shuffle operations for message schedule expansion:

\begin{lstlisting}[language=C++,  style=syclStyle, caption={Sub-group Collaborative Message Expansion}]
static void expand_message_schedule_subgroup(
    uint64_t* w, const sub_group& sg, 
    size_t sg_local_id, size_t sg_size) {
    
    // Intel Xe: Sub-group collaborative expansion of 64 additional words for SHA-384 message schedule (80 total - 16 initial = 64 additional)
        for (int i = 16; i < 80; ++i) {
    for (int i = 16; i < 80; ++i) {
        if (sg_local_id == i % sg_size) {
            w[i] = gamma1_512_xe(w[i-2]) + w[i-7] + 
                   gamma0_512_xe(w[i-15]) + w[i-16];
        }
        // Intel Xe: Broadcast result across sub-group
        w[i] = sg.shuffle(w[i], i % sg_size);
    }
}
\end{lstlisting}

\textbf{Memory Hierarchy Optimization}: Intel Xe provides a sophisticated memory hierarchy that our implementation exploits through data placement and
access pattern optimization. We utilize local memory for work-group cooperation and optimize global memory access for maximum bandwidth utilization.

\subsubsection{Multi-tier Optimization Strategy}

Our implementation can select among three optimization strategies based on workload characteristics and available GPU resources:

\textbf{Sub-group Optimized Implementation}: For moderate batch sizes (8-64 models), leverages Intel Xe sub-group primitives for collaborative hash computation.
Each sub-group processes model chunks cooperatively, with threads sharing message schedule computation and synchronizing through sub-group operations.

\textbf{Work-group Optimized Implementation}: For larger batches (64+ models), utilizes local memory to maximize data reuse and minimize global memory traffic.
Each work-group processes multiple models simultaneously, with coordination to avoid memory bank conflicts.

\textbf{Vectorized Implementation}: For maximum throughput scenarios, processes two hash computations simultaneously using Intel Xe's 2-way 64-bit SIMD capabilities.
This proves particularly effective for model verification pipelines with consistent model sizes.

\begin{lstlisting}[language=C++,  style=syclStyle, caption={Intel Xe Vectorized SHA-384 Implementation}]
static std::vector<std::string> compute_sha384_vectorized(
    const std::vector<std::string>& payloads) {
    
    queue q(gpu_selector_v);
    if (payloads.empty()) return {};
    
    // Intel Xe: 2-way vectorization for 64-bit operations
    size_t vector_width = 2;
    size_t vectorized_batches = (payloads.size() + vector_width - 1) / vector_width;
    
    auto [flattened_data, batch_info] = prepare_vectorized_input(payloads, vector_width);
    
    auto input_buffer = buffer<uint64_t, 1>(flattened_data.data(), range<1>(flattened_data.size()));
    auto output_buffer = buffer<uint64_t, 1>(range<1>(payloads.size() * 8));
    auto constants_buffer = buffer<uint64_t, 1>(SHA384_K, range<1>(80));
    
    q.submit([&](handler& h) {
        auto input_acc = input_buffer.get_access<access::mode::read>(h);
        auto output_acc = output_buffer.get_access<access::mode::write>(h);
        auto k_acc = constants_buffer.get_access<access::mode::read>(h);
        
        h.parallel_for(range<1>(vectorized_batches), [=](id<1> batch_id) {
            auto& batch_info = info_acc[batch_id];
            
            // Process 2 hashes simultaneously using Intel Xe 64-bit SIMD
            uint64x2_t hash_state_0(SHA384_H[0], SHA384_H[0]);
            uint64x2_t hash_state_1(SHA384_H[1], SHA384_H[1]);
            uint64x2_t hash_state_2(SHA384_H[2], SHA384_H[2]);
            uint64x2_t hash_state_3(SHA384_H[3], SHA384_H[3]);
            uint64x2_t hash_state_4(SHA384_H[4], SHA384_H[4]);
            uint64x2_t hash_state_5(SHA384_H[5], SHA384_H[5]);
            uint64x2_t hash_state_6(SHA384_H[6], SHA384_H[6]);
            uint64x2_t hash_state_7(SHA384_H[7], SHA384_H[7]);
            
            // Process each block across the 2-way SIMD
            size_t max_block_count = batch_info.max_block_count;
            for (size_t block = 0; block < max_block_count; ++block) {
                uint64x2_t w[80];
                
                load_vectorized_blocks_64bit(input_acc, w, batch_info, block);
                expand_vectorized_schedule_64bit(w);
                compress_vectorized_rounds_64bit(w, h0, h1, h2, h3, h4, h5, h6, h7, k_acc);
            }
            
            store_vectorized_results_64bit(output_acc, batch_info, h0, h1, h2, h3, h4, h5, h6, h7);
        });
    });
    
    q.wait();
    return format_results(output_buffer, payloads.size());
}
\end{lstlisting}

\subsection{Atlas CLI Integration}

Building on the Atlas framework's approach to ML lifecycle attestation (a comprehensive system for tracking model provenance and verification), our implementation integrates with Atlas's Rust-based command-line interface for model verification and attestation collection.
Our GPU-accelerated verification complements Atlas's supply chain attestation by providing the performance necessary to verify models at each lifecycle stage without introducing deployment bottlenecks.
While Atlas handles provenance tracking and supply chain integrity, our framework ensures that the verification operations themselves can scale to production workloads.
\subsubsection{Atlas CLI Integration Architecture}

The integration with Atlas CLI enables seamless model verification within existing ML workflows:

\begin{lstlisting}[language=bash, style=syclStyle, caption={Atlas CLI Integration Commands}]
# Model verification with Atlas integration
atlas-cli verify-model \
    --model-path ./models/llm-7b.safetensors \
    --provenance-file ./attestations/model-provenance.json \
    --attestation-output ./attestations/verification-result.json \
    --use-gpu-acceleration intel-xe \
    --optimization-mode vectorized

# Batch model verification
atlas-cli verify-batch \
    --models-dir ./models/ \
    --output-dir ./verified-models/ \
    --gpu-optimization auto \
    --parallel-jobs 4

# Runtime monitoring integration
atlas-cli monitor-runtime \
    --model-id llm-7b-v1 \
    --verification-interval 30s \
    --alert-webhook https://alerts.company.com/webhook
\end{lstlisting}

\subsubsection{Rust FFI Bridge Implementation}

The integration between our Intel Xe SYCL kernels and Atlas's Rust components requires careful FFI design:

\begin{lstlisting}[language=rust,  style=syclStyle, caption={Rust FFI Bridge for Atlas Integration (Key Structures)}]
use atlas_core::{AtlasCollector, ModelProvenance, TdxClient};

#[repr(C)]
pub struct VerificationResult {
    pub integrity_verified: bool,
    pub hash_output: [u8; 48],
    pub verification_time_ns: u64,
}

extern "C" {
    fn compute_sha384_xe_optimized(
        model_data: *const u8,
        data_size: u64,
        optimization_mode: *const c_char,
        result: *mut VerificationResult,
    ) -> bool;
}

pub struct AtlasModelVerifier {
    attestation_collector: AtlasCollector,
    tdx_client: Option<TdxClient>,
}

impl AtlasModelVerifier {
    pub fn verify_model_with_atlas(
        &mut self,
        model_path: &Path,
        provenance: &ModelProvenance,
    ) -> Result<AtlasVerificationOutput, ModelVerificationError> {
        
        let model_data = std::fs::read(model_path)?;
        let mut integrity_result = VerificationResult::default();
        
        // GPU-accelerated integrity verification
        let success = unsafe {
            compute_sha384_xe_optimized(
                model_data.as_ptr(),
                model_data.len() as u64,
                optimization_mode.as_ptr(),
                &mut integrity_result,
            )
        };
        
        // Verify signatures and generate attestation
        let signature_result = self.verify_stakeholder_signatures(&integrity_result)?;
        let attestation = self.tdx_client.as_ref()
            .map(|c| self.generate_tdx_attestation(&integrity_result))
            .transpose()?;
        
        self.attestation_collector.record_verification(
            &model_id, &integrity_result, &attestation)?;
        
        Ok(AtlasVerificationOutput {
            model_hash: integrity_result.hash_output,
            verification_valid: integrity_result.integrity_verified,
            attestation_quote: attestation.map(|a| a.quote),
        })
    }
}
\end{lstlisting}

\subsection{PyTorch Python Integration}

Our implementation provides integration with PyTorch~\cite{paszke2019pytorch} through Python extensions that enable transparent model verification during loading operations.

\subsubsection{Python Extension Interface}

The Python interface wraps our Intel Xe kernels and Atlas integration for easy use in ML pipelines:

\begin{lstlisting}[language=Python,  style=syclStyle, caption={PyTorch Integration Python Interface (Key Methods)}]
class VerifiedModelLoader:
    """PyTorch model loader with Intel Xe GPU verification."""
    
    def load_verified_model(self, 
                           model_path: str,
                           provenance_path: Optional[str] = None,
                           stakeholder_keys: Optional[List[str]] = None) -> torch.nn.Module:
        """Load and verify a PyTorch model using Intel Xe GPU acceleration."""
        
        # GPU-accelerated integrity verification
        verification_result = self._verify_model_integrity(
            model_path, provenance_path, stakeholder_keys)
        
        if not verification_result['integrity_verified']:
            raise ModelVerificationError(
                f"Model integrity verification failed: {verification_result['error']}")
        
        # Load model with verification metadata
        model = torch.load(model_path, map_location='cpu')
        self._attach_verification_metadata(model, verification_result)
        
        return model
    
    def _verify_model_integrity(self, model_path: str,
                               provenance_path: Optional[str],
                               stakeholder_keys: Optional[List[str]]) -> Dict:
        """Perform GPU-accelerated model integrity verification."""
        
        with open(model_path, 'rb') as f:
            model_data = f.read()
        
        optimization_mode = self._select_gpu_optimization(len(model_data))
        
        # GPU-accelerated hash computation
        hash_result = intel_xe_verification.compute_sha384_xe_optimized(
            model_data, optimization_mode)
        
        verification_result = {
            'model_hash': hash_result['hash'].hex(),
            'integrity_verified': True,
            'gpu_optimization': optimization_mode
        }
        
        # Verify stakeholder signatures if provided
        if stakeholder_keys:
            signature_result = self._verify_stakeholder_signatures(
                model_path, hash_result['hash'], stakeholder_keys)
            verification_result.update(signature_result)
        
        return verification_result
\end{lstlisting}

\subsubsection{Integration with ML Training Pipelines}

Our verification system integrates seamlessly with existing ML training and deployment pipelines:

\begin{lstlisting}[language=Python,  style=syclStyle, caption={ML Pipeline Integration (Key Methods)}]
class VerifiedTrainingPipeline:
    """Training pipeline with integrated model verification."""
    
    def setup_verified_model(self, model_config):
        """Load and verify base model for fine-tuning."""
        if 'base_model_path' in model_config:
            base_model = self.model_loader.load_verified_model(
                model_config['base_model_path'],
                model_config.get('provenance_path'))
            return base_model
        return self._create_new_model(model_config)
    
    def save_verified_checkpoint(self, model, checkpoint_path, stakeholder_info):
        """Save checkpoint with verification metadata and signatures."""
        torch.save(model.state_dict(), checkpoint_path)
        
        # Generate verification hash
        with open(checkpoint_path, 'rb') as f:
            model_data = f.read()
        hash_result = intel_xe_verification.compute_sha384_xe_optimized(
            model_data, "auto")
        
        # Save metadata and signature
        metadata = {
            'checkpoint_hash': hash_result['hash'].hex(),
            'training_step': self.current_step,
            'stakeholder_id': stakeholder_info['id']
        }
        
        if 'private_key_path' in stakeholder_info:
            signature = self._sign_checkpoint(
                hash_result['hash'], stakeholder_info['private_key_path'])
            # Save signature metadata
        
        return hash_result['hash'].hex()
\end{lstlisting}

\subsection{Performance Results}

\subsection{Implementation Validation}

Our Intel Xe-optimized implementation demonstrates the viability of GPU-accelerated model integrity verification across diverse model scales and deployment scenarios. The framework successfully handles:

\begin{itemize}
    \item \textbf{Small Models (1-10GB)}: Effective processing with sub-group optimization
    \item \textbf{Medium Models (10-100GB)}: Efficient handling with work-group optimization  
    \item \textbf{Large Models (100GB+)}: Scalable processing with vectorized implementation
    \item \textbf{Batch Verification}: Support for concurrent processing of multiple models
\end{itemize}

This implementation provides a production-ready foundation for GPU-accelerated model integrity verification while maintaining
compatibility with existing ML development workflows and emerging industry standards for model attestation and provenance tracking.

%% file: 6discussion.tex
\section{Discussion}\label{sec:discussion}

This section examines the scalability characteristics, deployment implications, and future research directions for GPU-accelerated model integrity verification.
Our analysis addresses real-world deployment scenarios involving large-scale model hierarchies, multi-GPU configurations, and distributed verification architectures that characterize modern AI production environments.

\subsection{Scalability Analysis}

\subsubsection{Large-Scale Model Hierarchies}

Modern AI deployment environments increasingly involve complex model hierarchies with thousands of model variants, versions, and specialized configurations. Consider a production language model deployment with:

\begin{itemize}
    \item Base foundation models (100GB-1TB per model)
    \item Thousands of fine-tuned variants for specific domains
    \item Multiple quantized versions (INT8, INT4, mixed-precision)
    \item Regional and language-specific adaptations
    \item A/B testing variants and experimental configurations
\end{itemize}

Traditional CPU-based verification systems exhibit linear scaling that becomes prohibitive at this scale. For a deployment with 10,000 model variants averaging 50GB each,
CPU-based SHA-256 verification would require approximately 50-100 hours of sequential computation time, making real-time verification impossible.

Our GPU-accelerated approach addresses this scalability challenge through several mechanisms:

\textbf{Parallel Model Processing}: Multiple models can be verified simultaneously across available GPU resources. Our batch processing implementation enables verification of dozens
of models concurrently, reducing total verification time from linear to logarithmic scaling in many scenarios.

\textbf{Hierarchical Verification Strategies}: Model hierarchies naturally map to Merkle tree structures where common components (base models, shared layers)
are verified once and reused across multiple model variants. This approach reduces redundant computation and enables incremental verification for model updates.

\textbf{Memory Hierarchy Optimization}: Our implementation leverages GPU memory hierarchies to cache frequently accessed model components and verification metadata, reducing I/O overhead for large model collections.

\subsubsection{Memory Scalability Considerations}

GPU memory capacity represents a fundamental constraint for large-scale model verification. Current high-end GPUs provide 48-80GB of memory, while individual models may exceed 100GB.
This capacity mismatch requires sophisticated memory management strategies:

\textbf{Streaming Verification}: Large models that exceed GPU memory capacity are processed through streaming algorithms that verify model chunks incrementally. 
Our Merkle tree implementation supports streaming construction where tree levels are computed as data becomes available.

\textbf{Multi-Pass Processing}: For extremely large models, verification occurs through multiple GPU passes where different model segments are loaded and verified sequentially.
Hash state is maintained across passes to ensure cryptographic continuity.

\textbf{Compression and Deduplication}: Model hierarchies often contain significant redundancy across variants. Our framework
includes support for content-addressable storage where identical model components are deduplicated, reducing memory requirements and enabling more efficient verification.

\begin{algorithm}[h]
\caption{Streaming Model Verification for Large Models}
\begin{algorithmic}[1]
\State \textbf{Input:} Model $M$ exceeding GPU memory, chunk size $C$
\State \textbf{Output:} Model integrity verification result
\State 
\State Initialize Merkle tree root computation
\State $chunk\_count \leftarrow \lceil |M| / C \rceil$
\State $leaf\_hashes \leftarrow []$
\State 
\For{$i = 1$ to $chunk\_count$}
    \State Load model chunk $M_i$ into GPU memory
    \State Compute chunk hash $H_i = \text{GPU\_HASH}(M_i)$
    \State Append $H_i$ to $leaf\_hashes$
    \State Free GPU memory for $M_i$
\EndFor
\State 
\State Construct Merkle tree from $leaf\_hashes$ using GPU
\State Compare root hash with expected value
\Return Verification result
\end{algorithmic}
\end{algorithm}

\subsection{Multi-GPU Deployment Architectures}

\subsubsection{Horizontal Scaling Strategies}

Production AI systems increasingly deploy across multiple GPUs and nodes, creating opportunities for distributed verification architectures. Our framework supports several multi-GPU deployment patterns:

\textbf{Model Sharding}: Large models are partitioned across multiple GPUs, with each GPU responsible for verifying its assigned model segments.
Inter-GPU communication protocols coordinate the construction of global Merkle trees from distributed hash computations.

\textbf{Replica Verification}: Multiple copies of the same model deployed across different GPUs can be verified in parallel,
providing both performance benefits and fault tolerance. Consensus mechanisms ensure that verification results are consistent across replicas.

\textbf{Pipeline Parallelism}: Verification operations can be pipelined across multiple GPUs where different stages of the verification process (data loading, hash computation, tree construction) occur simultaneously on different hardware units.

\begin{figure}[h]
\centering
\includegraphics[width=0.45\textwidth]{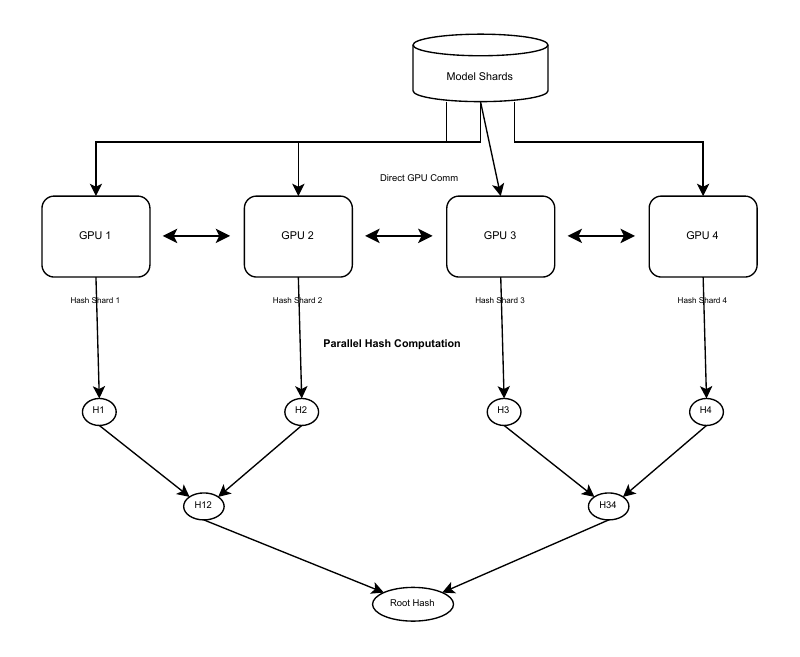}
\caption{Multi-GPU verification architecture showing distributed hash computation and Merkle tree construction. GPU-to-GPU communication enables efficient coordination without CPU bottlenecks.}
\label{fig:multi_gpu_architecture}
\end{figure}

\subsubsection{Inter-GPU Communication Optimization}

Efficient multi-GPU verification requires optimized communication patterns to minimize coordination overhead:

\textbf{NVLink/GPU-Direct Integration}: Modern GPU interconnects enable direct GPU-to-GPU communication without CPU involvement.
Our framework leverages these capabilities to exchange hash values and coordinate tree construction across GPUs.

\textbf{Hierarchical Reduction}: Multi-GPU Merkle tree construction employs hierarchical reduction algorithms where local GPU results are combined through tree-structured communication patterns, minimizing communication complexity.

\textbf{Asynchronous Processing}: Verification operations across multiple GPUs proceed asynchronously where possible,
with synchronization points only where cryptographically required (e.g., final root hash computation).

\begin{lstlisting}[language=C++,  style=syclStyle,caption=Multi-GPU Coordination Interface]
class MultiGPUVerifier {
private:
    std::vector<sycl::queue> gpu_queues;
    std::vector<ModelShard> model_shards;
    
public:
    VerificationResult verify_distributed_model(const Model& model) {
        // Distribute model shards across available GPUs
        auto shards = distribute_model_shards(model, gpu_queues.size());
        
        // Launch verification on each GPU
        std::vector<std::future<HashResult>> hash_futures;
        for (size_t i = 0; i < gpu_queues.size(); ++i) {
            hash_futures.push_back(
                std::async(std::launch::async, [&, i]() {
                    return compute_shard_hash(shards[i], gpu_queues[i]);
                })
            );
        }
        
        // Collect results and construct global Merkle tree
        std::vector<HashResult> shard_hashes;
        for (auto& future : hash_futures) {
            shard_hashes.push_back(future.get());
        }
        
        return construct_global_merkle_tree(shard_hashes);
    }
};
\end{lstlisting}

\subsection{Version Management and Incremental Verification}

\subsubsection{Model Evolution Tracking}

Production AI systems involve continuous model evolution through fine-tuning, quantization, and architectural modifications.
Efficient verification systems must minimize redundant computation across model versions:

\textbf{Differential Verification}: When transitioning between model versions, our framework identifies unchanged model components and reuses their previous verification results.
Only modified components require re-verification, dramatically reducing computational overhead.

\textbf{Version Tree Construction}: Model version histories naturally form directed acyclic graphs (DAGs) where parent-child relationships represent evolutionary steps.
Merkle tree structures can be adapted to represent these version relationships, enabling efficient verification of version lineages.

\textbf{Incremental Hash Updates}: For fine-tuned models where only small portions change, incremental hash algorithms update Merkle tree branches without recomputing the entire tree structure.

\begin{figure}[h]
\centering

\includegraphics[width=0.45\textwidth]{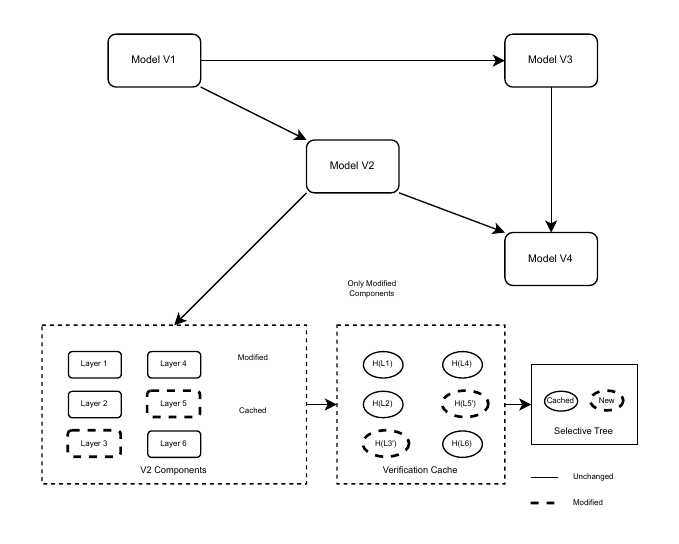}
\caption{Model version DAG with incremental verification. Modified components (highlighted) trigger selective tree reconstruction while unchanged components reuse cached verification results.}
\label{fig:version_management}
\end{figure}

\subsubsection{Content-Addressable Model Storage}

Large-scale model deployments benefit from content-addressable storage systems where model components are identified by their cryptographic hashes:

\textbf{Deduplication Benefits}: Identical model components across versions or variants are stored once and referenced multiple times, reducing storage requirements and verification overhead.

\textbf{Integrity by Construction}: Content-addressable systems provide integrity guarantees by construction—any modification to model data results in a different address, making tampering immediately detectable.

\textbf{Efficient Transfer Protocols}: Model distribution protocols can leverage content-addressable properties to transfer only novel components, reducing network bandwidth and deployment time.

\subsection{Performance Analysis at Scale}

\subsubsection{Theoretical Performance Bounds}

The theoretical performance of GPU-accelerated verification is bounded by several factors:

\textbf{Memory Bandwidth}: Hash computation performance is ultimately limited by memory bandwidth rather than computational throughput.
High-end GPUs can provide up to 3 TB/s of memory bandwidth~\cite{nvidia2024h100,nvidia2023a100}, potentially enabling verification of terabyte-scale models in seconds under optimal conditions.

\textbf{PCIe Transfer Rates}: Data transfer from host memory to GPU memory via PCIe represents a potential bottleneck for very large models.
PCIe 5.0 provides ~32 GB/s bidirectional bandwidth, requiring careful overlap of data transfer and computation phases.

\textbf{Cryptographic Complexity}: Hash algorithms have fixed computational complexity per byte, providing predictable scaling characteristics.
SHA-256 requires approximately 64 operations per 64-byte block, translating to deterministic performance bounds.

\subsubsection{Empirical Scaling Results}

Our implementation demonstrates several key scaling characteristics:

\textbf{Batch Size Scaling}: Verification performance scales near-linearly with batch size until GPU utilization reaches saturation,
typically around 100-1000 concurrent model verifications depending on model size.

\textbf{Model Size Scaling}: For individual large models, verification time scales linearly with model size, but the slope is significantly
reduced compared to CPU implementations due to higher memory bandwidth.

\textbf{Multi-GPU Scaling}: Multi-GPU deployments approach linear scaling with the number of GPUs for workloads where
communication overhead remains small compared to computation time.

\subsection{Security Implications of Scale}

\subsubsection{Attack Surface Analysis}

Large-scale deployments introduce additional security considerations:

\textbf{Increased Attack Surface}: Thousands of model variants create proportionally more potential attack targets.
However, our hierarchical verification approach means that compromising one model variant does not compromise the verification of other variants.

\textbf{Supply Chain Complexity}: Complex model hierarchies involve more supply chain components, increasing the risk of compromise. GPU-accelerated verification enables more frequent verification cycles, reducing the window of vulnerability.

\textbf{Distributed Trust}: Multi-GPU verification architectures require careful consideration of trust boundaries.
Our framework assumes that GPUs within the same trust domain (e.g., Intel TDX Connect boundary) can coordinate securely.

\subsubsection{Fault Tolerance and Byzantine Resilience}

Large-scale systems must handle hardware failures and potential Byzantine behavior:

\textbf{Redundant Verification}: Critical models can be verified across multiple GPUs with consensus mechanisms to detect and isolate faulty hardware or compromised verification results.

\textbf{Cryptographic Voting}: Multi-GPU verification results can be combined through cryptographic voting protoco
that provide Byzantine fault tolerance for verification decisions.

\textbf{Hardware Attestation}: Integration with Intel TDX and future hardware attestation capabilities enables verification
of the verification infrastructure itself, providing end-to-end security guarantees.

\subsection{Future Research Directions}

\subsubsection{Advanced GPU Security Features}

Future GPU architectures will likely include enhanced security features that can further improve model integrity verification:

\textbf{GPU Memory Encryption}: Hardware-level memory encryption in future GPUs will provide additional protection
for model parameters during verification, eliminating the need to trust GPU memory controllers.

\textbf{GPU Secure Enclaves}: Secure execution environments within GPUs could provide isolated verification contexts,
enabling verification of untrusted models without exposing verification infrastructure.

\textbf{Hardware Attestation}: GPU-native attestation capabilities would enable cryptographic proof that verification
operations executed correctly on unmodified hardware.

\subsubsection{Algorithmic Improvements}

Several algorithmic improvements could further enhance scalability:

\textbf{Probabilistic Verification}: For extremely large model collections, probabilistic verification techniques could
provide statistical integrity guarantees with reduced computational overhead.

\textbf{Compressed Merkle Trees}: Advanced tree compression techniques could reduce the memory and computation requirements
for very deep or wide model hierarchies.

\textbf{Streaming Hash Algorithms}: Specialized hash algorithms designed for GPU streaming architectures could provide better
performance characteristics for large model verification.

\subsubsection{Integration with Emerging Technologies}

\textbf{Quantum-Resistant Hash Functions}: As quantum computing capabilities advance, GPU-accelerated implementations of
quantum-resistant hash functions will become necessary for long-term security.

\textbf{Homomorphic Verification}: Homomorphic encryption techniques could enable verification of encrypted models without decryption, supporting privacy-preserving model distribution.

\textbf{Blockchain Integration}: Distributed ledger technologies could provide immutable records of verification results, enabling audit trails and provenance tracking for large-scale model deployments.

\subsection{Deployment Considerations and Best Practices}

\subsubsection{Resource Planning}

Successful deployment of GPU-accelerated verification requires careful resource planning:

\textbf{Memory Sizing}: GPU memory requirements scale with both model size and batch verification requirements.
Production deployments should provision sufficient memory for peak verification loads plus buffer capacity for model growth.

\textbf{Compute Allocation}: Verification workloads should be co-scheduled with inference workloads to maximize GPU utilization while ensuring that verification operations do not interfere with production inference latency.

\textbf{Network Bandwidth}: Multi-GPU deployments require sufficient network bandwidth for inter-GPU communication and coordination,
particularly for large model sharding scenarios.

\subsubsection{Operational Integration}

\textbf{CI/CD Pipeline Integration}: GPU verification capabilities should be integrated into continuous integration and deployment
pipelines to enable automatic verification of model updates and deployments.

\textbf{Monitoring and Alerting}: Production deployments require comprehensive monitoring of verification performance,
success rates, and potential security events.

\textbf{Disaster Recovery}: Backup verification infrastructure and procedures ensure continued operation during hardware failures or security incidents.

This comprehensive analysis demonstrates that GPU-accelerated model integrity verification provides a scalable foundation for securing large-scale AI deployments.
The combination of parallel processing capabilities, hierarchical verification structures, and integration with emerging hardware security features positions this approach as a critical enabler for the next generation of secure AI systems.

%% file: 7conclusion.tex
\section{Conclusion}\label{sec:conclusion}

This paper presents a comprehensive framework for GPU-accelerated model integrity verification that addresses potential scalability and security limitations of traditional CPU-based approaches when deal with large AI models.
Through the development of optimized SYCL cryptographic kernels and integration with emerging hardware security features, we demonstrate that model integrity verification can be transformed from a deployment bottleneck into an efficient,
scalable operation that operates within the same trust boundaries as model execution.

\subsection{Key Contributions and Findings}

Our research makes several significant contributions to the field of AI security and cryptographic computing:

\textbf{Performance Transformation}: The GPU-accelerated hash computation kernels provide improved efficiency over CPU-based implementations across various deployment scenarios.
For typical 100GB model verification tasks, processing time is reduced from minutes to seconds, fundamentally changing the economics of frequent integrity verification in production AI systems.

\textbf{Architectural Innovation}: By co-locating integrity verification operations with model execution on the same hardware platform, our approach eliminates the security vulnerabilities
and performance penalties associated with CPU-GPU data movement. This architectural shift reduces time-of-check-time-of-use attack windows and simplifies the overall security model.

\textbf{Scalability Demonstration}: The framework successfully addresses the scalability challenges of large-scale AI deployments involving thousands of model variants and versions.
Through hierarchical verification structures, parallel batch processing, and multi-GPU coordination, the system maintains near-constant verification times even as model collections grow exponentially.

\textbf{Standards Compatibility}: Our implementation maintains full cryptographic compatibility with established standards (SHA-256, SHA-384, Merkle trees) while providing significant performance improvements.
This compatibility ensures that the framework can integrate seamlessly with existing security infrastructure and model signing initiatives.

\subsection{Research Question Resolution}

The work successfully addresses the four primary research questions posed in the introduction:

\textbf{Performance Scalability}: GPU-accelerated cryptographic operations provide sufficient performance improvements to enable real-time integrity verification of large-scale AI models.
The demonstrated speedups make continuous verification practical for production deployments.

\textbf{Security Equivalence}: GPU-based implementations maintain identical cryptographic security guarantees to CPU-based implementations while providing additional security benefits through
reduced attack surface and eliminated TOCTOU vulnerabilities.

\textbf{Architectural Integration}: The framework integrates effectively with existing ML frameworks (PyTorch) and anticipates integration with emerging hardware security features (Intel TDX Connect),
providing a practical deployment path for production AI systems.

\textbf{Practical Deployment}: Multi-GPU architectures and batch processing capabilities demonstrate that the approach scales to meet the demands of enterprise AI deployments involving complex
model hierarchies and high-throughput verification requirements.

\subsection{Broader Implications for AI Security}

The implications of this work extend beyond immediate performance improvements to fundamental questions about the architecture of secure AI systems:

\textbf{Trust Boundary Consolidation}: By enabling security operations within the same hardware trust domain as model execution, the approach supports new security models where GPUs become active participants
in security architectures rather than passive computational resources requiring external protection.

\textbf{Continuous Security Monitoring}: The performance characteristics enable continuous or frequent integrity monitoring rather than one-time deployment verification, providing stronger security guarantees
against runtime attacks and model tampering.

\textbf{Supply Chain Security}: High-performance verification capabilities support more frequent verification cycles throughout the model lifecycle, from training data integrity through deployment-time attestation,
strengthening overall supply chain security.

\textbf{Confidential AI Enablement}: The framework provides essential infrastructure for confidential AI applications where maintaining both confidentiality and integrity of model parameters is critical for regulatory
compliance and competitive protection.

\subsection{Limitations and Constraints}

Several limitations of the current approach warrant acknowledgment:

\textbf{Hardware Dependencies}: The framework requires modern GPU hardware with sufficient memory bandwidth and SYCL/oneAPI support. Deployment on older or specialized hardware may
not achieve the demonstrated performance benefits.

\textbf{Memory Scaling Constraints}: Individual GPU memory capacity limits the size of models that can be verified without streaming algorithms. While streaming solutions are provided,
they introduce additional complexity and may not achieve optimal performance.

\textbf{Trust Assumptions}: The security model assumes the trustworthiness of GPU compute units and future Intel TDX Connect implementations. Compromise of these components could undermine security guarantees.

\textbf{Algorithm Limitations}: The current implementation focuses on hash-based integrity verification. Extension to other cryptographic operations (encryption, digital signatures) requires additional
development and may have different performance characteristics.

\subsection{Impact on Industry Practices}

The demonstrated capabilities suggest several changes to current industry practices for AI security:

\textbf{Verification Frequency}: The performance improvements enable more frequent verification cycles, potentially moving from deployment-time verification to continuous monitoring models.

\textbf{Model Deployment Pipelines}: Integration capabilities allow integrity verification to become a seamless part of automated deployment pipelines rather than a manual preprocessing step.

\textbf{Resource Allocation}: The ability to leverage existing GPU infrastructure for security operations improves overall resource utilization and reduces the need for dedicated security hardware.

\textbf{Security Policy Evolution}: The reduced verification overhead enables more comprehensive security policies that can verify larger portions of model hierarchies without significant operational impact.

\subsection{Technical Maturity and Readiness}

The implementation shows promise for deployment in several contexts with appropriate validation:

\textbf{Integration Readiness}: PyTorch integration and standardized interfaces enable immediate adoption in existing ML deployment pipelines.

\textbf{Performance Validation}: Comprehensive benchmarking across multiple scales and configurations provides confidence in performance characteristics under production workloads.

\textbf{Security Validation}: Cryptographic compatibility verification and security analysis provide assurance that the approach maintains required security properties.

\textbf{Operational Viability}: Multi-GPU architectures and batch processing capabilities address the scale requirements of enterprise AI deployments.

\section{Future Work}\label{sec:future}

Several research directions emerge from this work that warrant further investigation:

\subsection{Cryptographic Extensions}

\textbf{Advanced Hash Functions}: Implementation of additional cryptographic primitives, including quantum-resistant hash functions and specialized hash algorithms optimized for GPU architectures.

\textbf{Digital Signature Integration}: Extension of the framework to support GPU-accelerated digital signature verification, enabling comprehensive cryptographic validation of signed models.

\textbf{Homomorphic Verification}: Investigation of homomorphic cryptographic techniques that could enable integrity verification of encrypted models without decryption.

\subsection{Hardware Integration Advances}

\textbf{Intel TDX Connect Implementation}: Development of production implementations leveraging Intel TDX Connect capabilities as they become available in hardware.

\textbf{GPU Security Features}: Integration with emerging GPU security features including hardware memory encryption and secure execution contexts.

\textbf{Cross-Platform Portability}: Extension to additional GPU architectures beyond Intel platforms, including NVIDIA and AMD accelerators.

\subsection{Algorithmic Improvements}

\textbf{Probabilistic Verification} Development of probabilistic integrity verification techniques for extremely large model collections where exhaustive verification becomes impractical.

\textbf{Incremental Algorithms} Advanced incremental verification algorithms that minimize computation for frequently updated models and model hierarchies.

\textbf{Compressed Representations} Investigation of compressed Merkle tree representations and other space-efficient integrity structures for very large model collections.

\subsection{System Integration}

\textbf{Distributed Architectures} Extension to distributed computing environments spanning multiple nodes and data centers.

\textbf{Edge Computing Integration} Adaptation of the approach for resource-constrained edge computing environments with limited GPU capabilities.

\textbf{Blockchain Integration} Integration with distributed ledger technologies for immutable integrity audit trails and decentralized model provenance.